\documentclass[useAMS,usenatbib]{mn2e}
\usepackage{times}
\usepackage{graphicx}

\usepackage[T1]{fontenc}
\usepackage{aecompl}

\begin{document}

\title[Star Formation in a Time-Dependent LW Background]{High-Redshift Star Formation in a Time-Dependent Lyman-Werner Background}

\author[E. Visbal et al.]{Eli Visbal$^1$\thanks{visbal@astro.columbia.edu} \thanks{Columbia Prize Postdoctoral Fellow in the Natural Sciences}, Zolt\'{a}n Haiman$^1$, Bryan Terrazas$^1$, Greg L. Bryan$^1$, Rennan Barkana$^2$ \\ $^1$Department of Astronomy, Columbia University, 550 West 120th Street, New York, NY, 10027, U.S.A. \\
$^2$ Raymond and Beverly Sackler School of Physics and Astronomy, Tel Aviv University, Tel Aviv 69978, Israel}

\maketitle

\begin{abstract}
The first generation of stars produces a background of Lyman-Werner (LW) radiation which can photo-dissociate molecular hydrogen, increasing the mass of dark matter halos required to host star formation.  Previous studies have determined the critical mass required for efficient molecular cooling with a constant LW background.  However, the true background is expected to increase rapidly at early times.  Neglecting this evolution could underestimate star formation in small halos that may have started to cool in the past when the LW intensity was much lower.   Background evolution is a large source of uncertainty in pre-reionization predictions of the cosmological 21cm signal, which can be observed with future radio telescopes.  To address this, we perform zero-dimentional one-zone calculations that follow the density, chemical abundances, and temperature of gas in the central regions of dark matter halos, including hierarchical growth and an evolving LW background.  We begin by studying the physics of halos subjected to a background that increases exponentially with redshift.  We find that when the intensity increases more slowly than $J_{\rm LW}(z) \propto 10^{-z/5}$, cooling in the past is a relatively small effect.  We then self-consistently compute the cosmological LW background over $z=15-50$ and find that cooling in the past due to an evolving background has a modest impact.  Finally, we compare these results to three-dimensional hydrodynamical cosmological simulations with varying LW histories.  While only a small number of halos were simulated, the results are consistent with our one-zone calculations.
\end{abstract}


\section{Introduction}

The first stars in the universe formed during the period between recombination ($z\sim1100$) and cosmic reionization ($z\sim7-15$).  Fortunately, due to the lack of heavy elements during that epoch, modeling these stars is a relatively clean theoretical problem.  Primordial, metal-free (Pop III) stars are believed to form in small, $\sim 10^5 M_\odot$ dark matter halos that undergo efficient molecular hydrogen cooling  \citep{2002Sci...295...93A,1996ApJ...464..523H,1997ApJ...474....1T,2001ApJ...548..509M}.  While it will not be possible in the near future to detect individual objects at these extremely high redshifts, future measurements of the 21cm hydrogen line will have the ability to test theoretical predictions \citep[for reviews see][]{2006PhR...433..181F, 2012RPPh...75h6901P}.  Future observatories such as the Square Kilometer Array (SKA) \citep{2004NewAR..48.1029C}, sensitive to redshifts as high as $z \sim 30$, could infer the properties of early stars through their impact on the intergalactic medium (IGM).

An important aspect of Pop III star formation is negative feedback from the Lyman-Werner (LW) background  \citep{1997ApJ...476..458H}.  LW photons (11.2--13.6 eV) destroy molecular hydrogen and raise the minimum mass of dark matter halos where star formation can occur.  Previous works utilizing both hydrodynamical simulations \citep[e.g.][]{2001ApJ...548..509M,2007ApJ...671.1559W,2008ApJ...673...14O} and analytic calculations \citep[e.g.][]{1997ApJ...474....1T,2000ApJ...534...11H} have estimated the critical dark matter halo mass required for efficient molecular cooling as a function of background intensity and redshift, $M_{\rm crit}(J_{\rm LW},z)$.    To compute $M_{\rm crit}(J_{\rm LW},z)$, all of these studies made the simplifying assumption of a constant LW background intensity.  However, the strength of the background is expected to increase rapidly as a function of time.  Ignoring this time dependence could potentially underestimate the amount of star formation.  This is because a halo that starts to cool in the past, when the background was lower, may continue to cool even if the background increases such that its mass is below the critical value at a later time.  

The impact of time evolution in the LW background on early star formation is currently a large source of uncertainty in predicting the pre-reionization 21cm signal \citep{2013MNRAS.432.2909F,2012Natur.487...70V, 2012ApJ...760....3M}.  \cite{2013MNRAS.432.2909F} account for this by assuming that $M_{\rm crit}(J_{\rm LW},z)$ should be evaluated with the value of $J_{\rm LW}$ sometime in the past, when irreversible cooling presumably began.  A range of values for this time delay were selected based on the spherical collapse model, leading to significant changes in the 21cm power spectrum.  The appropriate choice of this delay is unclear \emph{ab initio}. 

In this paper, we perform zero-dimensional one-zone calculations of negative LW feedback with a time-evolving background in an effort to better understand the physics of primordial star formation.  Our calculations follow the density and chemistry of star-forming dark matter halos that grow hierarchically within a rapidly increasing LW background.  To track the growth of dark matter halos we generate Monte Carlo (MC) merger trees and follow the mass of the main progenitor (found by starting at the final halo and going back in time along the branch with the most-massive progenitor at each merger).  

Our analysis of these one-zone calculations is divided into two main parts.  First, we consider the impact of a time-dependent LW background on star formation in individual dark matter halos.  For a given halo mass, if and when cooling occurs depends on the particular merger history of each halo.  For this reason, rather than $M_{\rm crit}(J_{\rm LW},z)$, it is more precise to describe the cooling threshold in terms of the fraction of halos at a given mass and redshift that have cooled, $f_{\rm cool}(z,M,J_{\rm LW}(z))$.  Here $J_{\rm LW}(z)$ represents the history of the background at all redshifts. We compute $f_{\rm cool}$ at $z=25$ for a range of exponentially increasing $J_{\rm LW}(z)$.  Cooling in the past due to a lower LW background is a relatively small effect for backgrounds that increase more slowly than $J_{\rm LW} \propto 10^{-z/5}$.  In the second portion of our analysis, we self-consistently compute the cosmological $J_{\rm LW}(z)$ across all relevant redshifts.  Because the LW background increases sufficiently slowly, we find that cooling depends mainly on the current LW intensity and not on its history.  We compare these results to three-dimensional cosmological simulations, including hydrodynamics and primordial chemistry, and find that they give the same main conclusion, the intensity of the LW background in the past does not have a strong impact on when a halo can start to cool.

This paper is structured as follows.  Is $\S2$ we analyze the impact of a time-evolving LW background on individual dark matter halos.  We describe our computational method and prescription for generating MC merger trees in the first two subsections.   In $\S$2.3, we demonstrate how $f_{\rm cool}$ changes for different LW background histories.  In $\S$3, we use our method to self-consistently compute the background strength over all relevant redshifts.  In $\S$4, we present results from hydrodynamic cosmological simulations with evolving LW backgrounds.  We discuss our results and conclusions in $\S$5.  Except for the three-dimensional numerical simulations, we assume a $\Lambda$CDM cosmology consistent with the latest constraints from Planck \citep{2013arXiv1303.5076P}: $\Omega_\Lambda=0.68$, $\Omega_{\rm m}=0.32$, $\Omega_{\rm b}=0.049$, $h=0.67$, $\sigma_8=0.83$, and $n_{\rm s} = 0.96$.  The slightly different cosmology used in the numerical simulations should not have a significant impact on our results.

\section{Molecular Hydrogen Cooling in a Time-Evolving Lyman-Werner Background}
\subsection{Method}
We first focus on the physics of molecular hydrogen cooling in individual dark matter halos.
Previous studies have calculated the critical halo mass, $M_{\rm crit}(J_{\rm LW},z)$, above which efficient molecular cooling will occur for a fixed LW background \citep{2001ApJ...548..509M,2007ApJ...671.1559W,2000ApJ...534...11H}.  As discussed above, approximating the LW intensity as constant in time can underestimate the number of halos that cool.  For a rapidly increasing LW background, some halos below $M_{\rm crit}(J_{\rm LW},z)$ may have started to cool in the past when the background was much lower.  When cooling first occurs depends both on the evolution of the LW background and each halo's individual assembly history.  Thus, instead of a critical mass, the cooling threshold for star formation is better described by the fraction of halos which cooled at any time in the past for a specific background history, $f_{\rm cool}(z,M,J_{\rm LW}(z))$.  Here $f_{\rm cool}$ is a functional (rather than a simple function) since it depends on $J_{\rm LW}(z)$ at all previous times.

We compute $f_{\rm cool}$ with one-zone calculations of the chemical evolution in the central regions of dark matter halos similar to \cite{2001ApJ...546..635O}.  These include both the density evolution due to hierarchical structure formation and a time varying LW background.  We track the abundance of nine different chemical species: H, H$^-$, H$^+$, He, He$^-$, He$^+$, He$^{++}$, H$_2$, H$_2^+$, and $e^-$.  The initial abundances are assumed to equal the post-recombination values in the intergalactic medium \citep{1996ApJ...460..556A} and the reaction rates from \cite{2010MNRAS.402.1249S} are used.  The time dependent LW background is assumed to have a spectral energy dependence of $J_{\rm LW} \propto \nu^{-1}$.  Throughout this paper we use $J_{\rm LW}(z)$ to denote the value at 13.6 eV.

\begin{figure} 
\includegraphics[width=84mm]{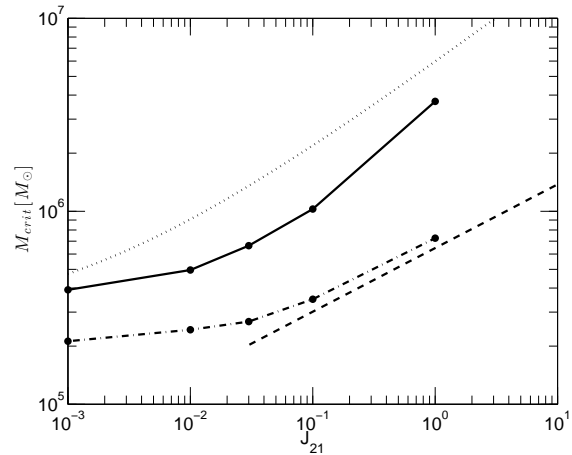}
\caption{The critical mass for efficient molecular hydrogen cooling with a constant LW background at $z=25$.  The dashed-dotted line is computed with our one-zone chemical network computation described in $\S$2.1 for $c_1=c_2=c_3=1$ and the dashed curve is the analytic estimate given by Eqn.~\ref{analytic_crit}.  There is good agreement between our calculations and the analytic approximation.  The solid line is for our model with $c_1=0.72$, $c_2=0.4$ and $c_3=0.25$. This is much closer to Eqn.~\ref{Mcrit_sims} (dotted curve), which matches results from detailed numerical simulations.  The LW background, $J_{21}$, is in units of ${\rm 10^{-21} ergs~s^{-1}~cm^{-2}~Hz^{-1}~sr^{-1}}$ and $M_{\rm crit}$ is in units of $M_\odot$.  
 } 
\label{tcool_check}
\end{figure}

To follow the density evolution of dark matter halos, we create MC merger trees and follow the main progenitor as a function of redshift (see $\S$2.2 for details).  As a halo grows, we assume that the gas in the central region is heated to the virial temperature of the main progenitor, $T_{\rm vir}$.  We expect this to be a good approximation before cooling becomes efficient \citep[see fig. 2 of][]{2001ApJ...548..509M}. In particular, we set the temperature to $T=T_{\rm vir}$ whenever the main progenitor increases its mass by 1 per cent or more.  In between these jumps, the temperature is computed from the reactions in the one-zone calculation.  Our results are not sensitive to the exact choice of how often we update the virial temperature.  We assume a central gas density equal to
\begin{equation}
\label{density_eqn}
n_{\rm H}(z) \sim c_1 \bar{n}_{\rm H}\left ( \frac{T_{\rm vir}}{T_{\rm IGM}}\right )^{\frac{c_2}{\gamma-1}} \sim 170  \times  c_1 \Omega_b h^2  \left ( \frac{T_{\rm vir}}{\rm 1000 K} \right )^{3c_2/2} {\rm cm^{-3}}.
\end{equation}
Here $\bar{n}_{\rm H}$ is the universal mean hydrogen density, $T_{\rm IGM} \sim 0.0135(1+z)^2$K is the temperature of the adiabatically cooling IGM, $\gamma=5/3$ is the adiabatic index for monoatomic gas, and $c_1$ and $c_2$ are free parameters we include to better match the critical cooling mass found in previous, detailed numerical simulations \citep{2001ApJ...548..509M,2007ApJ...671.1559W,2008ApJ...673...14O} (see Fig.~\ref{tcool_check}).    For $c_1=c_2=1$, this equation corresponds to the maximum density possible through adiabatic compression.

The value of the ${\rm H_2}$ column density, which determines the level of self-shielding from the LW background, is assumed to be $N_{\rm H_2} = \frac{1}{2}  r_{\rm J}  n_{\rm H_2}$, where $r_{J}=c_{\rm s} \sqrt{\pi/(G\rho)}$ is the Jeans length.  We set the central dark matter density (which is needed only to compute $r_{\rm J}$) to $10^4$ times the universal mean matter density at each $z$.  This corresponds to the average density within $10$ percent of the virial radius for an NFW profile with concentration $c = 4$ \citep{1997ApJ...490..493N}. We use the mean density within 10 per cent of the virial radius because this corresponds roughly to the size of constant density gas cores found in simulations before cooling occurs \citep[see e.g.][]{2014MNRAS.442L.100V}.  We find that our results are not highly sensitive to the exact density assumed.  For example, at $z=30$ with $J_{\rm LW}=0.1$, an order of magnitude increase or decrease in the density only changes the effective cooling mass by roughly 50 per cent.
We account for self-shielding by reducing the LW background by a factor $f_{\rm sh}(N_{\rm H_2},T)$ from \cite{2011MNRAS.418..838W} (see their eqn.~12).
    
We follow the chemical abundances, temperature, and density of the main progenitor over cosmic time and assume it cools and forms stars if the cooling time, $t_{\rm cool}$,
at some point in the assembly history is less than a fraction of the current Hubble time, $c_3$, chosen to match the results of previous simulations (i.e. a halo cools and forms stars if $t_{\rm cool}(z) < c_3 t_{\rm H}(z)$ at any previous $z$).  The cooling time is given by
\begin{equation}
\label{cool_eqn}
t_{\rm cool} = \frac{1.5n_{\rm g} k_{\rm B} T}{\Lambda n_{\rm H} n_{\rm H_2}},
\end{equation}
where $\Lambda$ is the ${\rm H_2}$ cooling function. 
Similar criteria have been used previously to study the constant LW background case \citep[e.g.][]{1997ApJ...474....1T}.  

Before exploring the effects of a time-varying LW background we test our model for the case of a constant background.  We compare our results to the analytic approximation for the critical cooling mass given by eqn.~25 of \cite{2001ApJ...548..509M}
\begin{equation}
\label{analytic_crit}
\frac{T{\rm crit}}{1000 {\rm K}}  \sim 0.36 \left ( (\Omega_{\rm b}h^2)^{-1} \left (4 \pi J_{\rm LW} \right)  \left( \frac{1+z}{20}\right )^{3/2}\right )^{0.22},
\end{equation}
where $T_{\rm crit}$ is the virial temperature corresponding to the critical mass and $J_{\rm LW}$ is in units of $10^{-21} {\rm ergs~s^{-1}~cm^{-2}~Hz^{-1}~sr^{-1}}$.  This formula is obtained by assuming H$_2$ photo-dissociation equilibrium and the density, temperature, and cooling assumptions described above for $c_1=c_2=c_3=1$.   In Fig.~\ref{tcool_check}, we plot the critical mass  of our model for $c_1=c_2=c_3=1$ (taken to be the mass where $f_{\rm cool}=0.5$) and the critical mass implied by this analytic expression at $z=25$.  There is very good agreement.  This analytic estimate is significantly lower than results from hydrodynamical simulations \citep{2001ApJ...548..509M,2007ApJ...671.1559W,2008ApJ...673...14O}, which find critical masses well approximated by
\begin{equation}
\label{Mcrit_sims}
M_{\rm crit} = 2.5 \times 10^5 \left ( \frac{1+z}{26} \right )^{-1.5}  \left (1+6.96 (4\pi J_{\rm LW})^{0.47} \right ).
\end{equation}
For the majority of our analysis, we adopt the parametrization $c_1=0.72$, $c_2=0.4$, and $c_3=0.25$.  This gives agreement in $M_{\rm crit}$ with Eqn. \ref{Mcrit_sims} to within a factor of two (see Fig.~\ref{tcool_check}).  Additionally, this parameterization gives central densities that match those from the simulations of \cite{2009MNRAS.399.1650M} and \cite{2001ApJ...548..509M} to within a factor of a few.   We note that the purpose of this study is not to exactly determine the mass at which halos can cool for a given intensity of the LW background.  Our main goal is to determine whether or not the cooling threshold computed with an evolving background is different than that computed with a constant background.  We find that the self-consistently computed $J_{\rm LW}(z)$, described below, for both parametrizations of our model shown in Fig.~\ref{tcool_check} is not strongly affected by LW background evolution changing the cooling mass threshold.  This suggests that despite the small discrepancy between the cooling criterion of our one-zone model and three-dimensional simulations, our main conclusions are correct. Future work comparing the details  of one-zone models and hydrodynamic simulations (e.g. $n_{\rm H}(z)$ and the cooling criterion) will be required to eliminate this discrepancy, but this is beyond the scope of the present work.

\subsection{Monte Carlo merger trees}
\begin{figure} 
\includegraphics[width=84mm]{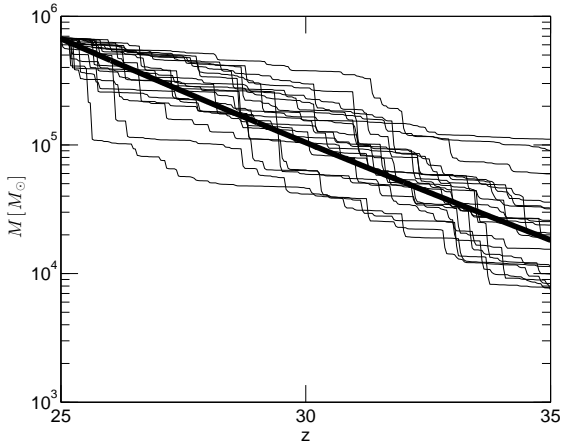}
\caption{Evolution of the main progenitor computed with the prescription described in $\S$ 2.2 for a representative subset of 10 dark matter halos with $M= 6.7 \times 10^{5} M_\odot$ at $z=25$.  The thick line is the analytic prediction of \citet{2014MNRAS.440...50M} (given by their eqn. 7).  } \label{trees}
\end{figure}

To compute the assembly history of dark matter halos, we generate  MC merger trees following the prescription of \cite{1993MNRAS.262..627L}.  We assume binary mergers going back in time with redshift steps $\Delta z$ apart.  At each redshift step, the first progenitor mass is determined randomly according to the spherical extended Press-Schechter (EPS) mass-weighted progenitor mass function and the second is set such that the sum of the progenitors equals the descendant halo mass.  If the smaller of the halo progenitors is less massive than the chosen mass resolution, $M_{\rm res}$, there is no merger at that time step.  We determine the random progenitor mass by inverting
\begin{equation}
x = \mathrm{erf} \left[  (\omega(z+\Delta z) - \omega(z)) / \sqrt{2 [ \sigma (M_1)^2 - \sigma(M_0)^2 ] } \right ].
\end{equation}
Here $x$ is a random number between 0 and 1, $ \omega(z) = \delta_{\rm crit}/D(z) \approx 1.686/D(z)$ is the critical over-density in the spherical collapse model extrapolated to the present day, $\sigma(M)$ is the root-mean-square density fluctuation on a scale corresponding to mass $M$, $M_0$ is the mass of the descendant halo, and $M_1$ is the mass of one of the two progenitors.  To follow the density of a growing halo we focus of the evolution of the main progenitor.   The main progenitor is  determined by starting at the final halo and going back in time along the branch of the most massive progenitor at each merger.  We adopt values of $\Delta z = 0.04$ and $M_{\rm res} = 10^3 M_{\odot}$.  We compute $\sigma(M)$ with the transfer functions of \cite{1998ApJ...496..605E}.  In Fig.~\ref{trees}, we plot the evolution of the main progenitor for merger trees originating at $z=25$.  We note that using the evolution of the most-massive progenitor across the entire tree (rather than just down the main branch) would not significantly alter our results.  We choose the main progenitor because jumping between different branches of the tree would not make sense in the context of following a halo's continuous change in chemistry.

To check the accuracy of this method we compare the mean and variance of the main progenitor to that computed with the technique of \cite{2008MNRAS.383..615N}, which agrees with the Millennium Run N-body simulation \citep{2005Natur.435..629S}.  We find excellent agreement, but note that we can only make a direct comparison with N-body simulations at much higher masses and lower redshifts than are applicable for high-redshift star formation.  In particular, halos in the Millennium Simulation simulation we compare to correspond to overdensities with $\sigma(M) > 1$, whereas the first star-forming halos we consider correspond to $\sigma(M) \approx 0.5$.  However, we are encouraged that our merger trees agree well with the analytic treatment of \cite{2014MNRAS.440...50M} for the masses and redshifts relevant to our study (see their eqn. 7, which is calibrated to the Millennium Simulation).

\subsection{Results}
To study the physics of molecular hydrogen cooling in the presence of an evolving LW background we perform calculations with a range of background histories parametrized by  
\begin{equation}
\label{test_bg}
J_{\rm LW}(z) = J_0 \times 10^{(z_0-z)/\alpha_{\rm LW}},  
\end{equation}
where $J_0$ and $\alpha_{\rm LW}$ are constant parameters and $z_0$ is the redshift of interest.  We focus on $z_0=25$, where LW feedback is expected to be important.  At much earlier times the background is negligible, while at much later times the background saturates and only atomic hydrogen cooling occurs.  This parametrization of time evolution is not meant to precisely match the cosmological background, but will help to elucidate the important physics.  In the next section, we self-consistently compute the background based on the expected abundance of sources at high redshifts.

\begin{figure} 
\includegraphics[width=84mm]{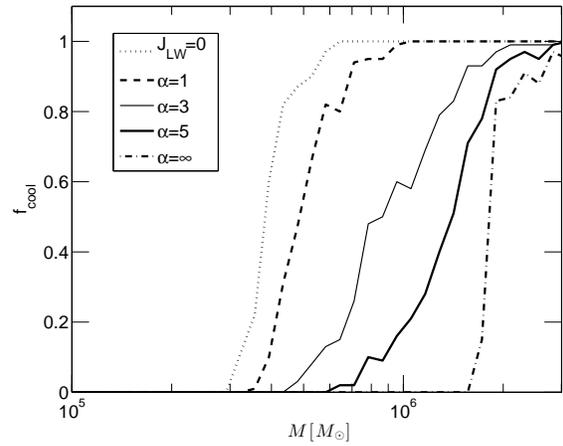}
\caption{The fraction of halos as a function of mass that have cooled by $z=25$ for a background given by Eqn.~\ref{test_bg}  with $J_0 = 0.3 \times10^{-21} {\rm ergs~s^{-1}~cm^{-2}~Hz^{-1}~sr^{-1}}$ and $z_0=25$.  The value of $\alpha_{\rm LW}$ is given by $\infty$, $5$, $3$, and $1$ for the  dot-dashed, thick-solid, thin-solid, and dashed lines, respectively.  The dotted line is for $J_0 = 0$.  The fraction in each mass bin is computed with 100 randomly generated dark matter halo merger histories.  As expected, more rapidly time-evolving backgrounds lead to star formation in smaller dark matter halos.} \label{cool_frac}
\end{figure}

In Fig.~\ref{cool_frac}, we show the fraction of halos at $z_0=25$ that have cooled as a function of halo mass for $J_0=0.3 \times 10^{-21} {\rm ergs~s^{-1}~cm^{-2}~Hz^{-1}~sr^{-1}}$ and $\alpha_{\rm LW }=1, 3, 5$ and $\infty$ according to the criterion in Eqn.~\ref{cool_eqn}.  As expected, more rapidly time-evolving backgrounds permit star formation in smaller dark matter halos.  For $\alpha_{\rm LW} > 5$, there is not a strong effect from the time dependence.  Since the LW background is unlikely to evolve much faster than this in any small redshift range, this suggests that the time dependence effect may not have a very large impact on the actual cosmological background.  This will be explored in the following section.

\begin{figure} 
\includegraphics[width=84mm]{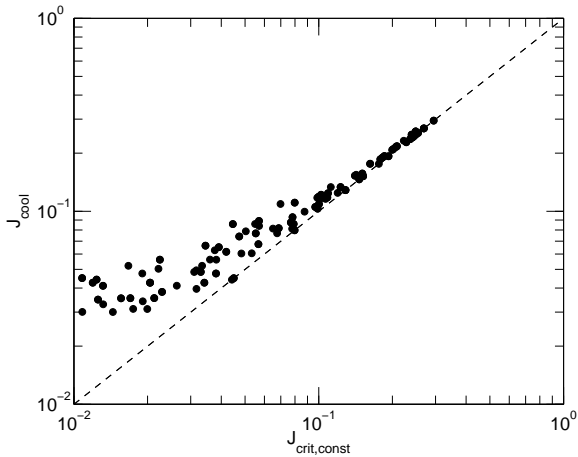}
\caption{$J_{\rm cool}$ versus $J_{\rm crit,const}(z)$ for a sample of halos that cool between $z=25-30$ when subjected to a background parametrized by $J_0=0.3 \times 10^{-21} {\rm ergs~s^{-1}~cm^{-2}~Hz^{-1}~sr^{-1}}$ and $\alpha_{\rm LW }=5$.  $J_{\rm cool}$ is the LW background intensity at the time when a halo meets our cooling criterion and $J_{\rm crit,const}(z)$ is the critical intensity  below which cooling can occur for a constant LW background at the same redshift.  We find that halos generally do not cool if the background is more than $\sim0.05 \times 10^{-21} {\rm ergs~s^{-1}~cm^{-2}~Hz^{-1}~sr^{-1}}$ above $J_{\rm crit,const}$.  That these halos do not cool until the background falls roughly to $J_{\rm crit,const}(z)$ allows us to explain why cooling in the past is not very important for our $\alpha_{\rm LW}=5$ example, but is for $\alpha_{\rm LW}=1$ (see Fig.~\ref{explanation}). } \label{jcool_const}
\end{figure}

In an effort to explain why the effect of an evolving background history is very large for $\alpha_{\rm LW}=1$ but not for $\alpha_{\rm LW}=5$, we examine the cooling of individual halos.  Specifically, we consider halos subjected to a background with $J_0=0.3 \times 10^{-21} {\rm ergs~s^{-1}~cm^{-2}~Hz^{-1}~sr^{-1}}$ and $\alpha_{\rm LW }=5$ that cool between $z=25$ and $z=30$.  In Fig.~\ref{jcool_const}, we plot the intensity of the time-dependent background at the redshift when each halo meets our cooling criterion, $J_{\rm cool}$, versus the critical level below which the halo would cool for a constant background at the same redshift, $J_{\rm crit,const}$.  With an increasing background, a halo will always cool if $J_{\rm LW}(z)$ falls below $J_{\rm crit,const}(z)$.  However, it is not clear \emph{a priori} how far above $J_{\rm crit,const}$ a time-varying background can be and still permit a halo to cool.  Fig.~\ref{jcool_const} illustrates that for $\alpha_{\rm LW}=5$, most halos will not cool if $J_{\rm LW}(z)$ is more than $\sim 0.05 \times 10^{-21} {\rm ergs~s^{-1}~cm^{-2}~Hz^{-1}~sr^{-1}}$ above $J_{\rm crit,const}$.

\begin{figure} 
\includegraphics[width=84mm]{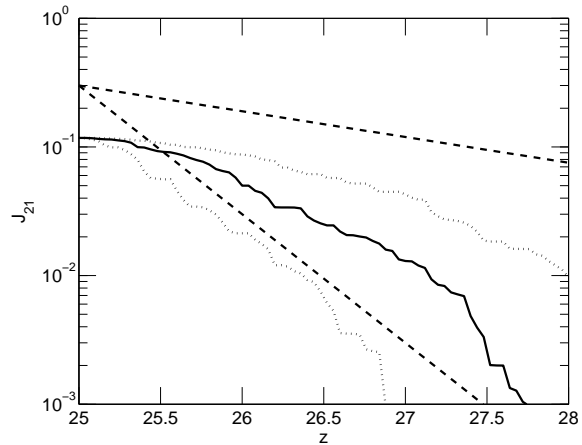}
\caption{The median critical intensity of a constant background, $J_{\rm crit}(M)$, for a sample of 100 halo merger histories with $M=10^6 M_\odot$ at $z=25$ as a function of redshift (solid curve).  The dotted lines denote the 25th and 75th $J_{\rm crit}(M)$ percentiles for the halo sample.  The thick dashed lines are $J_{\rm LW}(z)$ for $\alpha_{\rm LW }=5$ (upper curve) and $\alpha_{\rm LW }=1$ (lower curve).  Even though halos have a $J_{\rm crit}$ that is below the background at $z=25$, the $\alpha_{\rm LW }=1$ background falls below $J_{\rm crit}$ for most halos at higher redshift permitting cooling in the past.  On the other hand, the $\alpha_{\rm LW }=5$ background stays significantly above $J_{\rm crit}$ in the past which prevents cooling in most halos. }  \label{explanation}
\end{figure}

Given that halos in an $\alpha_{\rm LW}=5$ background do not cool until $J_{\rm LW}(z)$ is close to the critical intensity expected for a constant background, we can understand why the transition between evolving background history being an important effect and not having a large impact occurs between $\alpha_{\rm LW}=1$ and $\alpha_{\rm LW}=5$.  In Fig.~\ref{explanation}, we show how the critical intensity for a constant background, $J_{\rm crit}(M)$, evolves for the progenitors of a representative sample of $10^6 M_\odot$ halos at $z=25$.  We compute $J_{\rm crit}$ by interpolating the solid curve in Fig.~\ref{tcool_check} (we have ignored the redshift dependence of $J_{\rm crit}$ as it is relatively unimportant across the small range in $z$ considered here).  We note that this does not correspond exactly to critical LW background for halos with different merger histories, but due to the steepness of $f_{\rm cool}$ with mass for a constant background (see the dot-dashed curve in Fig.~\ref{cool_frac}), there is approximately a one-to-one correspondence between halo mass and $J_{\rm crit}$.  We compare $J_{\rm crit}$ as a function of redshift for our sample halos to $\alpha_{\rm LW}=1$ and $\alpha_{\rm LW}=5$ backgrounds.  At $z=25$, both backgrounds are significantly higher than $J_{\rm crit}$.  However, the $\alpha_{\rm LW}=1$ background falls below $J_{\rm crit}$ at higher redshift, permitting cooling in the past.  On the other hand, the $\alpha_{\rm LW}=5$ background remains significantly above $J_{\rm crit}$ illustrating why background history does not have a large impact on star formation for $J_{\rm LW}(z)$ that evolve this slowly.

\section{Self-Consistent Computation of the Lyman-Werner Background}
In the previous section, we analyzed the impact of a time-evolving LW background on star formation by considering a range of background parametrizations that decrease exponentially with redshift.  We found that for a background which increases less rapidly than $\alpha_{\rm LW}=5$, the time-evolution of the LW background does not have a large impact on star formation.

We now move beyond our artificial parametrization of the LW background and use our model to self-consistently compute the cosmological $J_{\rm LW}(z)$.  To accomplish this we start at very high redshift, where the LW background will have no impact, and take small steps down in redshift, computing the mean background intensity at each step.  We make a simple ``screening" assumption that LW photons travel through the intergalactic medium undisturbed until they redshift into a Lyman series line, at which point they are absorbed and removed from the LW band.  Thus, the maximum redshift, $z_{\rm m}$, a LW photon can be seen from at redshift $z$ is given by
\begin{equation}
\frac{1+z_{\rm m}}{1+z} = \frac{\nu_i}{\nu_{\rm obs}},
\end{equation}
where $\nu_{\rm obs}$ is the observed frequency and $\nu_i$ is the first Lyman line above this frequency.  We simplify this further by assuming that all LW radiation is seen out to
\begin{equation}
\frac{1+z_{\rm m}}{1+z} = 1.04,
\end{equation}
where 4 percent is roughly the amount that a typical photon in the LW band can redshift before hitting a Lyman line.  In reality, the intergalactic medium will cause a modest frequency dependent attenuation of the background with a characteristic ``sawtooth" shape \citep{1997ApJ...476..458H}.  However, we expect our simple screening assumption to be reasonably accurate, giving a factor of $\sim 10$ decrease compared to $z_{\rm m}=\infty$.  

The total background at each redshift step is given by
\begin{equation}
J_{\rm LW}(z) = \frac{c}{4\pi} \int_z^{z_{\rm m}} dz' \frac{dt_{\rm H}}{dz'} \left ( 1+z \right )^3 \epsilon(z'),
\end{equation}
where $c$ is the speed of light and $\epsilon(z)$ is the LW luminosity per frequency per comoving volume at redshift $z$.  To calculate $\epsilon(z)$, we assume that in each dark matter halo that undergoes cooling, a fraction of its baryons, $f_*$, is turned into stars over a period equal to the current cosmic time.  This gives
\begin{equation}
\epsilon(z) = \int_0^\infty dM f_*  f_{\rm cool}(M,z) \frac{dn}{dM}  \frac{\Omega_{\rm b}}{\Omega_{\rm m}} \frac{M}{m_{\rm p}} \left ( \frac{N_{\rm LW} E_{\rm LW}}{\Delta \nu_{\rm LW}} \right ) t_{\rm H}^{-1},
\end{equation}
where $dn/dM$ is the halo mass function \citep{1999MNRAS.308..119S}, $m_p$ is the proton mass, $N_{\rm LW}$ is the number of LW photons produced per baryon in stars, $E_{\rm LW}$ is the typical energy of a LW photon, and $\Delta \nu_{\rm LW}$ is the frequency range of the LW band.  We compute $f_{\rm cool}$ using the model described in $\S$2  with the $J_{\rm LW}(z)$ computed at previous redshift steps.   

\begin{figure} 
\includegraphics[width=84mm]{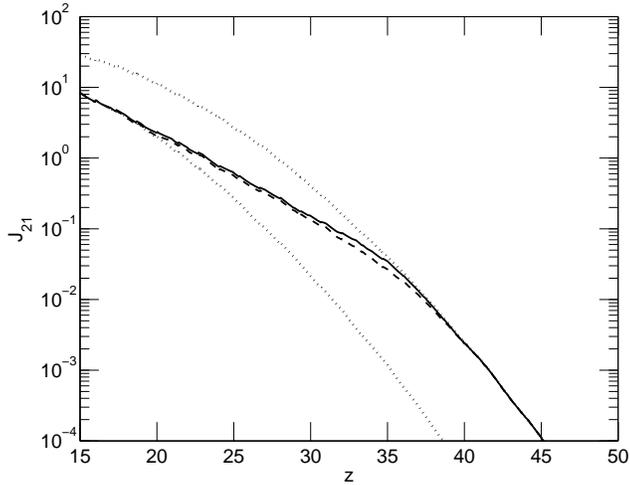}
\caption{The self-consistently computed $J_{\rm LW}(z)$ including (solid curve) and not including (dashed line) the effect of a time dependent LW background (in units of $10^{-21} {\rm ergs~s^{-1}~cm^{-2}~Hz^{-1}~sr^{-1}}$).  The dotted curves are the background produced with no LW feedback (top curve) and without molecular cooling (bottom curve).  The background does not decrease fast enough with redshift for its time-evolution to have a large impact on $f_{\rm cool}$. } \label{sc_plot}
\end{figure}

\begin{figure} 
\includegraphics[width=84mm]{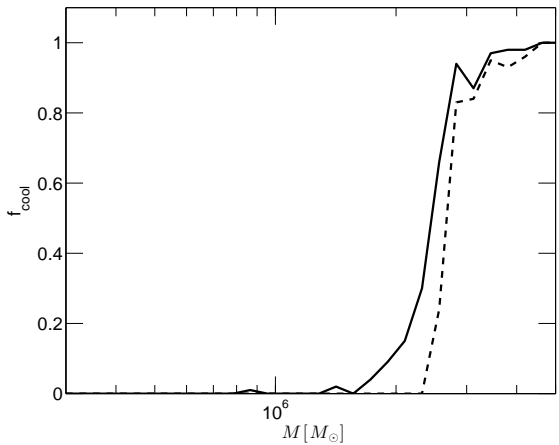}
\caption{The cooling fraction versus mass at $z=25$ calculated using the cosmologically self-consistent $J_{\rm LW}(z)$ (solid curve) and a constant background with $J_{\rm LW}(z=25) = 0.6 \times 10^{-21} {\rm ergs~s^{-1}~cm^{-2}~Hz^{-1}~sr^{-1}}$ (dashed curve).  The impact of background evolution is small. } \label{sc_fcool}
\end{figure}

In Fig.~\ref{sc_plot}, we plot the self-consistently computed $J_{\rm LW}(z)$ for assumed values of $f_* = 0.1$, $N_{\rm LW}=3400$, $E_{\rm LW}=1.9 \times 10^{-11}$~ergs and $\Delta \nu_{\rm LW}=5.8 \times 10^{14}$~Hz.  For comparison we also plot the background when the time dependence of the LW background is not taken into account.  In this case, $f_{\rm cool}(M,z)$ is computed with a background fixed at the value computed in the previous small redshift step.  We see that time dependence makes a relatively small difference (no more that $30$ percent at any redshift).   In Fig.~\ref{sc_fcool}, we compare $f_{\rm cool}(M)$ at $z=25$ computed with the cosmologically self-consistent background to that calculated assuming a constant background with the same intensity at that redshift.  There is very little effect from the time evolution (a $\sim 10$ percent shift in $f_{\rm cool}$).   This is consistent with the analysis presented in $\S$2, since over any small redshift range $J_{\rm LW}$ does not increase faster than an exponential function (Eqn.~\ref{test_bg}) with $\alpha_{\rm LW} \sim 5$  (except at very early times when the background is too low to have an effect).  We repeated this analysis for our one-zone model parametrized by $c_1 = c_2 = c_3 = 1$ and find that background evolution has a similarly small affect.  Our results are also qualitatively the same for different values of the star formation efficiency between $f_*=0.01-0.2$. 

\section{Numerical Simulations}

In order to double-check the merger-tree calculations just described, which can model many halos in a wide parameter space, but do not include the full three-dimensional structure of the merging halos, we also carry out a set of numerical simulations with an evolving LW background.  We use the adaptive mesh refinement (AMR) code Enzo \citep{Bryan1997, OShea2004, Enzo2013} and a nine species non-equilibrium chemical network (H, H$^+$, He, He$^+$, He$^{++}$, H$^-$, H$_2$, H$^+_2$, and e$^{-}$) that includes cooling \citep{Abel1997, Anninos1997}.  For these simulation, we adopt a WMAP7 compatible cosmological model with parameters  given by  $\Omega_{\Lambda,0} = 0.721$, $\Omega_{m,0} = 0.279$, $\Omega_b = 0.046$, $\sigma_8 = 0.817$, $n_s = 0.96$ and $h=0.701$ \citep{Komatsu2011}.  

To identify a region of interest, we carry out a low-resolution run ($128^3$ cells and particles, with 5 levels of refinement) in a box of size 1 Mpc, starting at $z=99$ and stopping at $z=15$.  We then repeat the calculation, focusing on a $0.25^3$ Mpc sub-region that hosts a range of halos including the largest halo in the entire box at $z=15$.  We do this by generating initial conditions in three additional (static) levels so that the effective initial resolution is $1024^3$ and the dark matter particle mass is 30 M$_\odot$.  During the run, adaptive refinement is included; we use up to 13 levels of refinement to ensure that the Jeans length is always resolved by at least 4 cells, and that the dark matter (baryonic) mass in a cell does not exceed 120 (15.5) M$_\odot$.  The gives a smallest cell size of 0.056 physical pc at $z=15.5$.  In addition, we include an artificial thermal pressure to ensure that even after we have reached the maximum level, the Jeans length did not drop below 10 cells \citep[as in][]{2001ApJ...548..509M}.

\begin{figure}
\begin{center}
\includegraphics[width=90mm]{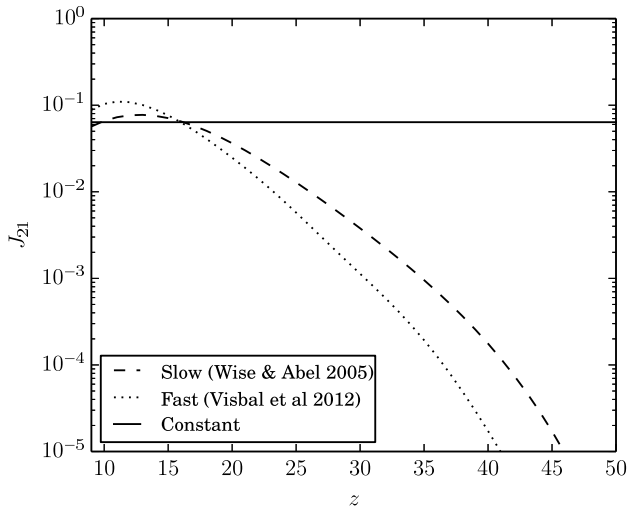}
\end{center}
\caption{The evolving LW background $J_{21}$ as a function of redshift for the three AMR simulations with a background discussed in this paper.  The slowly-varying profile is from \citet{WiseAbel2005} and the quickly varying profile is from \citet{2012Natur.487...70V}, although the amplitude of both have been adjusted to match at $z=15.5$.}
\label{sim_J21}
\end{figure}

To test the impact of a varying LW background, we repeat this calculation (with the same initial conditions) with four different configurations: (i) a constant LW background with $J_{21} = 0.06$, (ii) a slowing varying background taken from \citet{WiseAbel2005}, (iii) a more rapidly varying background from \citet{2012Natur.487...70V}, and (iv) a run with no LW background.  For all runs with a background, the amplitude of the LW fields were adjusted so that they all were the same at $z=15.5$.  The time evolving backgrounds are shown in Fig.~\ref{sim_J21}.

\begin{figure}
\begin{center}
\includegraphics[width=90mm]{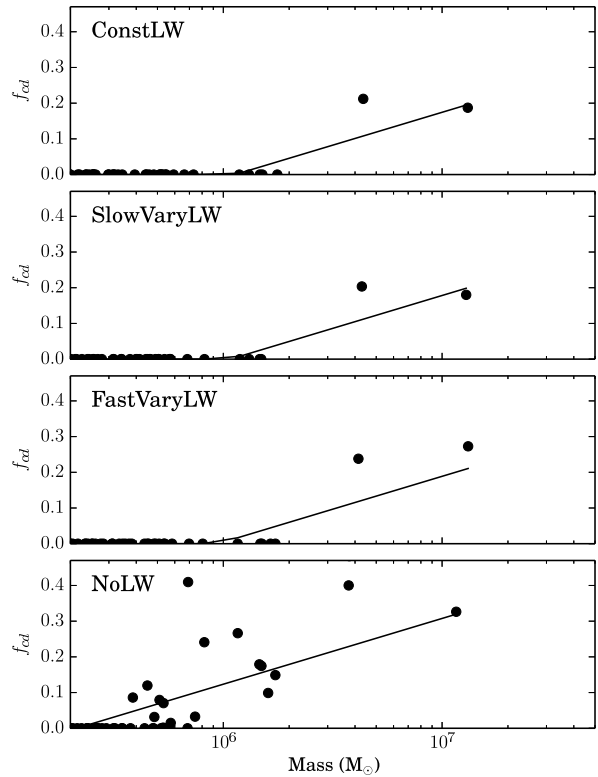}
\end{center}
\caption{The fraction of cold and dense gas ($T < 0.5 T_{\rm vir}$ and $n > 330$ cm$^{-3}$) at $z=15.5$ in each halo as a function of halo mass for the four numerical simulations discussed in this paper.  The solid line is a simple fitting formula as discussed in the text.}
\label{sim_fcd}
\end{figure}

To analyze the ability of the halos to cool by the end of the simulations at $z=15.5$, when the LW flux is identical, we follow the procedure described in \citet{2001ApJ...548..509M} -- in particular, we define the fraction of cold and dense gas ($f_{\rm cd}$) in a halo as the gas mass with $T < 0.5 T_{\rm vir}$ and $\rho > 10^{19}$ M$_\odot$ Mpc$^{-3} \sim 300$ cm$^{-3}$.  The halo finder {\it hop} \citep{hop} is used to identify halos and for each halo, we compute $f_{\rm cd}$, with the results shown in Fig.~\ref{sim_fcd}.  

In all three simulations with a background, only the two most massive halos are able to cool, regardless of how quickly the background varies.  Only for the run without any background (bottom panel), are significantly more halos able to cool.  We quantify this by fitting a simple expression $f_{\rm cd} = 0.08 \ln{(M/M_{\rm TH})}$, where $M_{\rm TH}$ is a measure of the threshold mass for cooling.  For the FastVary/SlowVary/Const runs we find only a small variation: $M_{\rm TH} = 0.95 \times 10^{6}$/ $1.08 \times 10^{6}$/ $1.1 \times 10^{6}$ M$_\odot$, while for the simulation without a LW background, we obtain  $M_{\rm TH} = 0.2 \times 10^{6}$ M$_\odot$.  For the two runs which can be compared with \citet{2001ApJ...548..509M} (ConstLW and NoLW), we find $M_{\rm TH}$ values which are in good agreement.

The limited halo statistics of the simulations prevent us from making definite statements, however, the results are certainly consistent with the merger tree findings described earlier, and do indicate that any effect from an evolving LW background appears to be slight.

\section{Conclusions}
Negative feedback from the LW background can destroy molecular hydrogen and prevent star formation in the high-redshift universe.  Previous studies calculating the critical dark matter halo mass required to host star formation have all considered a constant LW background.  This can underestimate the amount of star formation if cooling is able to occur in the past when the background was much lower.  This effect has been a major source of uncertainty in semi-numerical predictions of the pre-reionization 21cm signal \citep{2013MNRAS.432.2909F,2012Natur.487...70V,2013arXiv1306.2354F}.

We have completed the first comprehensive analysis of negative LW feedback that considers a time-dependent background.  We performed one-zone calculations that follow the density, temperature, and chemistry of gas in the central regions of dark matter halos subjected to a rapidly increasing LW background.  To characterize the mass threshold for efficient molecular cooling we computed $f_{\rm cool}(z,M,J_{\rm LW}(z))$, the fraction of halos that cool sometime before redshift $z$.  In $\S$2, we determined $f_{\rm cool}$ for a range of artificial LW background histories that grow exponentially according to Eqn.~\ref{test_bg}.  We found that for backgrounds which increase more slowly than $\alpha_{\rm LW} \sim 5$, the effect of time dependence is relatively small (see Fig.~\ref{cool_frac}).  

In \S3, we self-consistently computed the cosmological background from $z=15-50$ including the effect of time-dependence.  We compared this with the background obtained when $f_{\rm cool}(z,M)$ is calculated assuming constant $J_{\rm LW}$ (set to the value at the previous redshift step).  We find that time-dependence has a relatively small impact ($<30$ percent).  We found similar results with both the $c_1=0.72$, $c_2=0.4$, and $c_3=0.25$ and $c_1=c_2=c_3=1$ parametrizations of our model.  Thus, even though our model does not exactly match the critical cooling mass taken from numerical simulations, our overall conclusion that mainly the current $J_{LW}$ and not its history impacts star formation, seems robust.  This is also supported by the cosmological simulations presented in \S4.

Our results suggest that semi-numerical simulations of the large-scale distribution of the first stars will not be changed greatly by the LW time-dependance effect.  However, we point out that (even in the case of a constant background) $f_{\rm cool}(M,z)$ provides a better description of the cooling threshold than a simple step function, because cooling depends not only on a halo's current state, but also on its individual detailed assembly history.  Using $f_{\rm cool}$ calibrated with hydrodynamical simulations, rather than a cut-off at $M_{\rm crit}$ could increase the accuracy of the predicted 21cm signal.

We have not performed a detailed comparison of our one-zone calculations and the numerical simulations.  In future work it will be beneficial to make detailed checks of our assumptions regarding the central density and cooling criterion. As our simulations followed only a limited number of halos, future larger simulations with more halos would help to confirm our results.  Finally, we note that throughout this analysis we have ignored the impact of large-scale baryon-dark matter streaming velocities \citep{2010PhRvD..82h3520T,2012MNRAS.424.1335F}.  The $J_{\rm LW}(z)$ we compute will only apply to large-scale ($\sim 100 {\rm Mpc}$) regions with low streaming velocity.  However, since the relative velocity effect tends to delay the star formation, it seems likely that our overall conclusions will apply to fast moving regions as well.  This should be checked in detail with future hydrodynamical simulations.  

\section*{Acknowledgements}
EV was supported by the Columbia Prize Postdoctoral Fellowship in the Natural Sciences.  ZH was supported by NASA grant NNX11AE05G.  GLB was supported by NSF grant 1008134 and NASA grant NNX12AH41G.  RB was supported by Israel Science Foundation grant 823/09.

\bibliography{H2_bib}
\end{document}